\documentstyle[12pt]{article}
\oddsidemargin--1mm
\begin{document}
\newcommand{\pl}{\partial}
\newcommand{\be}{\begin{equation}}
\newcommand{\ee}{\end{equation}}
\newcommand{\ba}{\begin{eqnarray}}
\newcommand{\ea}{\end{eqnarray}}
\newcommand{\mbf}[1]{\mbox{\boldmath$ #1$}}
\def\<{\langle}
\def\>{\rangle}

\begin{center}
{\Large\bf On an effective action of monopoles in
Abelian projections of Yang-Mills theory}

\vskip 0.5cm
Sergei V. Shabanov {\footnote{on leave from Laboratory of
Theoretical Physics, JINR, Dubna, Russia.}}

\vskip 0.2cm

{\em Department of Mathematics, University of Florida,
Gainesville, FL-32611, USA.}
\end{center}

\begin{abstract}
A path integral description of an effective action of 
monopoles in Abelian projections of Yang-Mills theories
is discussed and used to establish a projection independence
of the effective action.
A dynamic regime in which the effective dynamics may contain
massive solitonic excitations is described.

\end{abstract}

Numerical simulations of  lattice Yang-Mills
theories show that certain topological defects, which occur upon
a partial gauge fixing \cite{2}, play an important role in the infrared
(nonperturbative) dynamics of Yang-Mills fields. A typical numerical
experiment of this kind would involve the following steps. Given
a set of Yang-Mills field configurations generated with the probability
$e^{-S_W}$, where $S_W$ is the Wilson action, one performs a gauge
transformation of
each configuration from the Wilson ensemble so that after the gauge
transformation the configuration
satisfies a certain gauge condition that breaks the original gauge
group to its maximal Abelian subgroup. The most popular gauge is the 
so called maximal Abelian gauge which is achieved by minimizing the
$L_2$ norm of non-Abelian components of Yang-Mills connections     
by means of suitable gauge transformations \cite{map}. Clearly,
this procedure would leave the maximal Abelian subgroup of the gauge
group unbroken. After the gauge fixing (or the Abelian projection),
the non-Abelian components are removed from every configuration
and only the Abelian components
of the gauge-fixed Wilson ensemble are used to compute expectation
values of some gauge invariant operators like, for instance, the Wilson
loop. An interesting feature of such a numerical experiment is that,
despite a substantial reduction of the degrees of freedom (removing
the non-Abelian components of connections), 
the string tension obtained from the Abelian (gauge-fixed) ensemble
is 92\% of the full string tension (computed in the original
Wilson ensemble) \cite{ad}. This is known as the Abelian dominance. 

Since this phenomenon does not occur if non-Abelian components are
removed {\em before} the gauge fixing, it is clear
that some relevant degrees of freedom have been transferred from
non-Abelian components to the Abelian ones upon the gauge transformation
which has been used to impose the gauge in the Wilson ensemble.
Hence, the effective Abelian theory cannot be a usual Maxwell theory.
The topology of the gauge group and its maximal Abelian subgroup
are different. Therefore any gauge fixing which breaks SU(N) to U(1)${}^{N-1}$
would have singularities. In particular, there are connections for which
the gauge transformation that minimizes
the $L_2$ norm of their non-Abelian components cannot
be regular everywhere in spacetime. It is easy to show that 
after the gauge transformation the Abelian components of such
connections would contain Dirac monopoles as topological defects
\cite{2},\cite{map}. Furthermore one can separate numerically
monopole (singular) 
and photon (regular) parts of the Abelian fields and use only
the ``monopole'' ensemble to compute the expectation value of
the Wilson loop and the string tension. The ``monopole'' string
tension differs from the ``Abelian'' one only by 5\% \cite{md}.     
Thus, the topological defects play the major role in nonperturbative
Yang-Mills theory in the maximal Abelian gauge. There is a strong numerical
evidence that this occurs in other Abelian projection gauges \cite{oapg}.
This phenomenon is called the monopole dominance.

It is well known that the physical configuration space, the space
of connections modulo gauge transformations (the orbit space),
 has a non-Euclidean
geometry and topology. When computing the functional integral
over the physical configuration space, one usually uses some 
local coordinates on it. The coordinates on the orbit space
can be obtained, for example, by a gauge fixing. Thanks to
the nontrivial topology of gauge orbits, the coordinate system
does not exist globally. Therefore any description based on local 
coordinates  will always exhibit singularities.
Although a ``physical'' interpretation of the coordinate singularities 
depends on the choice of coordinates (or, frankly speaking, on the choice
of the gauge), they are inevitable in any coordinate description.
Moreover, they have to be taken into account when computing the functional
integral in order to obtain a correct (gauge invariant) spectrum of physical
excitations in the theory \cite{pr}. It should be stressed that, though the 
coordinate singularities do depend on the choice of local coordinates,
their inevitable existence is essentially due to a non-Euclidean
structure of the physical configuration (or phase) space 
which is gauge independent \cite{pr}. That is, the geometry of the 
orbit space reveal itself through singularities in any 
coordinate description of the dynamics on the orbit space.   

Consider a local function $\phi(A)$ of the Yang-Mills connection $A_\mu$
which transforms in the adjoint representation under gauge transformations:
$\phi(A)\rightarrow \Omega\phi(A)\Omega^\dagger$. An Abelian projection
can be made by the gauge transformation $A_\mu\rightarrow A_\mu^\Omega =
\Omega A_\mu\Omega^\dagger + i\Omega\pl_\mu\Omega^\dagger$ where
for every $A_\mu$ the gauge group element $\Omega$ 
is determined by the condition
that $\phi(A^\Omega)=\Omega\phi(A)\Omega^\dagger$ is an element
of the Cartan algebra ($\Omega$ diagonalizes $\phi(A)$ 
in a matrix representation).
In other words, we impose the gauge condition that off-diagonal
components of $\phi(A)$ should be zero for every $A_\mu$. The Faddeev-Popov
determinant in this gauge is easy to compute $\Delta_{FP}[A]=
\prod_x\det({\rm ad}\, \phi(A))^2$, 
where the adjoint operator ${\rm ad}\,\phi$
acts on any element $y$ of the Lie algebra as ${\rm ad}\,\phi y=i[\phi,y]$. 
When computing the determinant, the zero modes of the operator 
${\rm ad}\,\phi$ associated with the residual Abelian gauge symmetry
must be removed. Note that the maximal Abelian subgroup of the gauge
group is isomorphic to the stationary group of $\phi$. 
One can show that the gauge group element $\Omega$ is singular whenever
$\det({\rm ad}\, \phi(A))^2$ vanishes at some points of spacetime \cite{s3}
and the Abelian components of the projected configuration $A_\mu^\Omega$
would contain Dirac monopoles at those points. Thus, if an Abelian 
projection gauge is used to establish 
local coordinates on the gauge orbit space, 
the ``monopole'' configurations would generally appear as singular points
of this coordinate system where the Faddeev-Popov determinant
vanishes.   

The results of the aforementioned 
numerical simulations show that the dynamics of Yang-Mills field
configurations, that look like Dirac magnetic monopoles {\em after} an
Abelian projection, captures essential features of
quantum Yang-Mills theory
at large distances (in the infrared limit).  A theoretical challenge
is to derive an effective theory for such degrees of freedom from the 
first principles.     
The conventional functional integral in the Abelian projection gauge
cannot be used as a starting point
because it is ill-defined ($\Delta_{FP}=0$) 
at relevant configurations. Our goal is to
develop the path integral formalism in which this problem is circumvented.

We begin with a trivial observation that   
the coordinate
singularities depend on the choice of a gauge and by changing the gauge
they can be moved away from a configuration
space region of interest. Suppose we have a general parameterization
of the Yang-Mills connections which,
upon an Abelian projection, become purely Abelian configurations
containing monopole-like topological
defects. An effective action of such configurations can be 
computed by the functional integral in a gauge in which the ``monopoles''
configurations are no longer coordinate singularities. A simple
analogy can be made with an ordinary integral over a sphere. If the origin
of the coordinate system is chosen to be on the north pole of the sphere,
then the south pole is the singular point (the point where two
geodesics through the origin intersect again). Suppose the integrand is such
that the stationary point approximation near the south pole gives a good
estimate of the integral. Clearly, to compute the integral,
it is more convenient to change the coordinates so that the singular      
point will be away from the south pole (e.g., by moving the origin to
the south pole).      
        
We shall elaborate this idea with an example of SU(2). A generalization
to the case of SU(N) is straightforward \cite{s2}.         
Let $\mbf{n}_0=(0,0,1)$ be a unit
isotopic vector. We are looking for connections $\mbf{A}_\mu$ which
can be transformed to configurations of the form
$\mbf{n}_0A_\mu$ by a suitable gauge
transformation. Clearly, such connections would have six functional
parameters, four in $A_\mu$ and the other two are associated with
parameters of gauge transformations from SU(2)/U(1), where 
the group U(1)$\sim$SO(2) is the stationary group of $\mbf{n}_0$.
The latter two
parameters are unified into a unit isotopic vector $\mbf{n}$, $\mbf{n}^2=1$.
By analyzing  generic (singular) 
gauge transformations of the connection $\mbf{n}_0A_\mu$,
we find that the connection \cite{cho,fn1,s1}
\be
\mbf{A}_\mu = \mbf{n}\times\pl_\mu\mbf{n} + \mbf{n} C_\mu
\label{1}
\ee
can be made purely Abelian by a suitable
gauge transformation, $\mbf{A}_\mu\rightarrow\mbf{n}_0(C_\mu +C_\mu^m)$,
where $C_\mu^m = \mbf{n}_0\cdot(\pl_\mu\mbf{\xi}\times\mbf{\xi})$       
and $\mbf{\xi}=(\sin(\theta/2)\cos\varphi,\sin(\theta/2)\sin\varphi,
\cos(\theta/2))$ if ${\bf n}=(\sin\theta\cos\varphi$,$\sin\theta$ $\sin\varphi$,
$\cos\theta)$. If we take, for example, $\mbf{n}=\mbf{x}/r$, $r^2=\mbf{x}^2$, 
the first term in (\ref{1}) will be a vector potential of the Wu-Yang monopole.
Upon the Abelian projection it produces an addition $C_\mu^m$ to the regular
Maxwell potential $C_\mu$ which is the vector potential of the Dirac magnetic
monopole localized at the origin and with the Dirac string extended along
a negative part of the $z$-axis. 
In a similar way one can also parameterize ``Abelian'' connections
in the SU(N) gauge theory \cite{fn2,s2} which would generate monopoles
upon an Abelian projection.

To develop an effective theory for the field $\mbf{n}$, one needs 
a change of variables in the space of connections. That is, six more
functional variables have to be added to the connection (\ref{1}):
\be          
\mbf{A}_\mu = \mbf{n}\times\pl_\mu\mbf{n} + \mbf{n} C_\mu + \mbf{W}_\mu\ ,
\label{2}
\ee
where $\mbf{W}_\mu\cdot \mbf{n}=0$ and $\mbf{W}_\mu$ contain only six
independent variables. There are infinitely many ways to parameterize
$\mbf{W}_\mu$ by six  variables. It can be done either 
implicitly \cite{s1} or explicitly \cite{fn2,s2}. 
In general, one can say that $\mbf{W}_\mu$ should satisfy two
more conditions
\be
\mbf{\chi}(\mbf{W},\mbf{n},C)=0\ ,\ \ \ \ \mbf{n}\cdot\mbf{\chi}\equiv 0\ .
\label{3}
\ee 
For example, one can choose \cite{s1}
\be 
\mbf{\chi}=\pl_\mu{\bf W}_\mu + C_\mu{\bf n}\times {\bf W}_\mu +
{\bf n} (\pl_\mu{\bf n}\cdot{\bf W}_\mu) =0\ .
\label{4}
\ee
The condition (\ref{4}) means that  $\nabla_\mu\mbf{W}_\mu =0$ where
the covariant derivative $\nabla_\mu$ is taken for the connection (\ref{1}).
An example of an explicit parameterization can be found if we unify
six independent components of $\mbf{W}_\mu$ into an antisymmetric tensor 
$W_{\mu\nu}=-W_{\nu\mu}$. Then one can set \cite{s2}
\be
\mbf{W}_\mu = W_{\mu\nu}\mbf{n}\times\pl_\nu\mbf{n}\ .
\label{5}
\ee
It is easy to find $\mbf{\chi}(\mbf{W}, \mbf{n})=0$ whose solution
is given by (\ref{5}). The analysis can be extended to the SU(N) case
\cite{fn2,s2}.

It is rather straightforward to show \cite{s1}
that the ``monopole'' configurations
of the Wilson ensemble in the maximal Abelian gauge are described 
by the vector potential $C_\mu^m(\mbf{n})$ introduced after Eq.(\ref{1})
if $\mbf{W}_\mu$ in the change of variables (\ref{2}) 
satisfies the condition (\ref{4}).
In general, for any Abelian projection one can find a parameterization
of $\mbf{W}_\mu$ such that the monopole defects are always described by
$C_\mu^m(\mbf{n})$ \cite{s2}. In the new variables (\ref{2}), an
Abelian projection is described by the gauge $\mbf{n}=\mbf{n}_0$ which
is singular if $C_\mu^m(\mbf{n})$ carries Dirac monopoles, i.e., 
this gauge does not exist everywhere in spacetime. Moreover the dynamics
favors configurations which become coordinate singularities in this gauge.
    
Since (\ref{2}) is a change of variables, the Wilson ensemble $\mbf{A}_\mu$ 
can be used to generate
the ensembles of $C_\mu[\mbf{A}]$, $\mbf{W}_\mu[\mbf{A}]$ and 
$\mbf{n}[\mbf{A}]$. The inverse transformation is nonlocal so $\mbf{n}$
is a nonlocal functional of $\mbf{A}_\mu$. In principle, the effective
action of $\mbf{n}$ can be computed numerically by means of the inverse
Monte-Carlo method  \cite{imc}. Given an ensemble of $\mbf{n}$ one
could try to find the probability which generates it. This method has
already been used to compute an effective action of the monopole current
in lattice gauge theories \cite{imcj}.

In a theoretical analysis, the change of variables (\ref{2}) in the 
functional integral allows one to avoid the singularities of the
Faddeev-Popov action in an Abelian projection gauge. The idea is
to choose a gauge so that the Faddeev-Popov determinant does not
vanish for the ``monopole'' configurations $\mbf{A}_\mu =\mbf{n}\times
\pl_\mu\mbf{n}$. To compute the Faddeev-Popov determinant, one has
to find the gauge transformation law in the new variables. 
An infinitesimal  gauge transformation 
of the SU(2) connection reads
\be
\delta \mbf{ A}_\mu = \nabla_\mu (\mbf{ A})\mbf{\varphi}=
\pl_\mu \mbf{\varphi} + \mbf{ A}_\mu \times \mbf{\varphi}
\ .
\label{3b}
\ee
From (\ref{2}) we infer
\be 
C_\mu = \mbf{n}\cdot \mbf{A}_\mu\  ,\ \ \ \
\mbf{W}_\mu = \mbf{n}\times \nabla_\mu (\mbf{A})\mbf{n}\ .
\label{3ab}
\ee
Substituting these relations into (\ref{3}) and solving them for 
$\mbf{n}$ (two equations for two independent variables in $\mbf{n}$),
we find $\mbf{n}=\mbf{n}(\mbf{A})$. The latter together with (\ref{3ab})
specifies the inverse change of variables. Let $\delta\mbf{n}$ be
an infinitesimal gauge transformation of $\mbf{n}$. Then from 
(\ref{3ab}) and (\ref{3b}) we deduce that
\ba
\delta C_\mu &=& \mbf{A}_\mu\cdot(\delta\mbf{n}-\mbf{n}\times\mbf{\varphi})
+ \mbf{n}\cdot \pl_\mu\mbf{\varphi}\ ,\label{4b}\\
\delta\mbf{W}_\mu &=&
\mbf{W}_\mu \times\mbf{\varphi} - \mbf{n}[\mbf{W}_\mu\cdot(\delta\mbf{n}-
\mbf{n}\times\mbf{\varphi})] + \mbf{n}\times \pl_\mu(
\delta\mbf{n}-\mbf{n}\times\mbf{\varphi})\ ,\label{5b}
\ea
where we have used that $\mbf{n}\cdot\delta\mbf{n}=0$.
An explicit form of $\delta\mbf{n}$ can be found 
from the equation $\delta\bar{\mbf{\chi}}(\mbf{n},\mbf{A})=0$
where $\bar{\mbf{\chi}}(\mbf{n},\mbf{A})$ is obtained by a substitution
of (\ref{3ab}) into $\mbf{\chi}(\mbf{W},C,\mbf{n})$. 
We emphasize that $\delta\mbf{n}$ is determined by the choice of
$\mbf{\chi}$ and so are $\delta C_\mu$ and $\delta
\mbf{W}_\mu$. 

Let us assume we have a gauge (e.g., a Lorentz gauge) 
in which the ``monopole'' configurations
$\mbf{n}\times\pl_\mu\mbf{n}$
are not coordinate singularities, that is, the Faddeev-Popov
determinant $\Delta_{FP}[\mbf{A}]$ does not
vanish for $\mbf{A}_\mu=\mbf{n}\times\pl_\mu\mbf{n}$. 
This would allow us to develop a perturbation theory
for computing an effective action of the field $\mbf{n}$.
After the change of variables (\ref{2}) in the integral
\be
{\cal Z} \sim \int D\mbf{A}_\mu \Delta_{FP}[\mbf{A}] e^{-S}\ ,
\label{6}
\ee
where $S$ is the Yang-Mills action with a gauge fixing term,
we can integrate out $C_\mu$ and $\mbf{W}_\mu$ by perturbation
theory. However, the Jacobian of the change of variables may
also be a source of singularities if it vanishes for the 
``monopole'' configurations. To compute the Jacobian, consider
the identity      
\be
1 =\int D\mbf{n}
\Delta[\mbf{A},\mbf{n}]\delta(\bar{\mbf{\chi}})\ ,\ \ \  
\Delta[\mbf{A},\mbf{n}] = \det [\delta\bar{\mbf{\chi}}/\delta\mbf{n}]\ .
\label{7}
\ee
Next, we insert the identity into (\ref{6}) and change the integration 
variables $\mbf{A}_\mu\rightarrow$
$\mbf{n}C_\mu + \mbf{W}_\mu$ with a {\em generic}
 $\mbf{W}_\mu$
perpendicular to $\mbf{n}$ so that $D\mbf{A}_\mu \sim
DC_\mu D\mbf{W}_\mu$. Finally, we shift 
the integration variables
$\mbf{W}_\mu \rightarrow \mbf{W}_\mu + \mbf{n}\times\pl_\mu\mbf{n}$
with the result \cite{s1}
\be
{\cal Z}\sim 
\int D\mbf{n} DC_\mu D\mbf{W}_\mu \Delta[\mbf{A},\mbf{n}]\Delta_{FP}[\mbf{A}]
\delta[\mbf{\chi}(\mbf{W},C,\mbf{n})]\, e^{-S}\ .
\label{t1}    
\ee
where $\mbf{A}_\mu$ is to be replaced by (\ref{2}). This completes
the path integral representation of Yang-Mills theory in the 
new variables (\ref{2}). The determinant $\Delta[\mbf{A},\mbf{n}]$
depends on the choice of $\mbf{\chi}$. The choice of $\mbf{\chi}$
must be such that $\Delta[\mbf{A},\mbf{n}]$ does not vanish for 
$\mbf{A}_\mu = \mbf{n}\times\pl_\mu\mbf{n}$.

Various
choices of $\mbf{\chi}$ can be regarded as {\em gauge fixing}
conditions for the gauge symmetry associated with a reparameterization
of $\mbf{W}_\mu$. Note that Eq.(\ref{2})
contains 14 functions in the right-hand side, while there are only 12
components in $\mbf{A}_\mu$. Therefore the gauge transformations
(\ref{3b}) would, in general, be induced by {\em five}-parametric
transformations of the new variables. There are two-parametric
transformations of the new variables under which $\mbf{A}_\mu$
remains invariant. Precisely this gauge freedom is fixed by (\ref{3})
and by the corresponding delta function in (\ref{t1}). 

The delta function in (\ref{t1}) can be put into the exponential
by means of the usual procedure of averaging over the gauge condition
with the result $S\rightarrow S + \int dx \mbf{\chi}^2/2$. The determinants
can also be lifted up to the exponential by introducing the ghost
variables. The effective action obtained in such a way will be invariant
under five-parametric BRST transformations. This symmetry can be used
to show that the effective action of the field $\mbf{n}$ does not depend
on the choice of $\mbf{\chi}$ in every order of perturbation theory \cite{s2}
for a rather large class of $\mbf{\chi}$'s that includes all Abelian
projections in which topological defects occur in the 
the Abelian components (i.e., they are monopoles).

It is noteworthy that one can give rather general arguments that
the gradient expansion of the effective action should have 
the form (in the leading order) \cite{fn1}  
\be
S_{eff} = \int dx\left\{ m^2 (\pl_\mu {\bf n})^2 + H_{\mu\nu}^2\right\}\ ,
\label{8}
\ee
where $H_{\mu\nu}={\bf n}\cdot
(\pl_\mu{\bf n}\times \pl_\nu{\bf n})$ and $m$ is a mass scale.
The field theory (\ref{8}) contains stable knot solitons. It might therefore
be possible to describe a nonperturbative (glueball) spectrum of Yang-Mills
theory by means of quantum theory of knot solitons \cite{f}. The mass scale
$m^2$ can be expressed as an expectation value of the antisymmeric field
$W_{\mu\nu}$ introduced in (\ref{5}) \cite{s2}
\be
\left\<\left(\pl_\mu W_{\nu\sigma}-\pl_\nu W_{\mu\sigma}\right)
\left(\pl_\mu W_{\nu\lambda}-\pl_\nu W_{\mu\lambda}\right)\right\>
\sim m^2 \delta_{\sigma\lambda}\ .
\label{9}
\ee
The above procedure of constructing a path integral for the  ``monopole''
degrees of freedom can be generalized to the SU(N) gauge group \cite{s2}.

\end{document}